# Secure Neighbor Discovery in Wireless Networks: Formal Investigation of Possibility

Marcin Poturalski, Panos Papadimitratos, Jean-Pierre Hubaux
Laboratory for Computer Communications and Applications
EPFL, Switzerland
{marcin.poturalski, panos.papadimitratos, jean-pierre.hubaux}@epfl.ch

## ABSTRACT

Wireless communication enables a broad spectrum of applications, ranging from commodity to tactical systems. *Neighbor discovery* (ND), that is, determining which devices are within direct radio communication, is a building block of network protocols and applications, and its vulnerability can severely compromise their functionalities. A number of proposals to *secure* ND have been published, but none have analyzed the problem formally. In this paper, we contribute such an analysis: We build a formal model capturing salient characteristics of wireless systems, most notably obstacles and interference, and we provide a specification of a basic variant of the ND problem. Then, we derive an *impossibility result* for a general class of protocols we term "time-based protocols," to which many of the schemes in the literature belong. We also identify the conditions under which the impossibility result is lifted. Moreover, we explore a second class of protocols we term "time- and location-based protocols," and prove they can secure ND.

## Categories and Subject Descriptors

C.2.0 [**Computer-Communication Networks**]: General—*Security and protection*

## General Terms

Security

## 1. INTRODUCTION

Wireless networking is a key enabler for mobile communication systems, that range from cellular infrastructure-based data networks and wireless local area networks (WLANs) to disaster-relief, tactical, and sensor networks, and short-range wire replacement and radio frequency identification (RFID) technologies. In all such systems, any two wireless devices communicate directly when in range, without the assistance of other devices. The ability to determine if direct, one-hop, communication takes place is fundamental. For example, a WLAN access point (AP) assigns a new IP address to a mobile station only when it is within the AP's coverage area. Or, a mobile node does not initiate a route discovery across a mobile ad hoc network (MANET) if a sought destination is already in its neighbor table. Or, an RFID tag will be read only if the signal transmitted by the tag can be received directly by the reader. These examples illustrate that, depending on whether another system entity, denoted as *node* in the rest of the paper, is a *neighbor* or not, actions are taken (e.g., by the AP or the router) or implications are derived (e.g., the RFID tag and reader are physically close). In other words, *discovering a neighbor*, or knowing that a node *is* a neighbor, is a common building block and enabler of diverse system functionality.

Nonetheless, if an attack against *neighbor discovery* (ND) can be perpetrated, such functionality can be abused. For example, letting legitimate nodes erroneously believe that they are neighbors allows the adversary to fully control communication across these artificial links. The threat lies in that the attacker can deny or derange communication at any point; this can happen exactly at the moment a message critical for the system operation is transmitted. In multi-hop networks, a "well-chosen" artificial link is likely to attract a considerable number of routes, with devastating effects: denial of communication across all these routes and significant disturbance in the flow of data. In a different scenario, misleading an RFID tag reader that the tag (and its owner) is physically close to the RFID reader, while this is not so, can enable the adversary to gain unauthorized access to the premises of the tag owner.

Such attacks against ND are easy to mount, because the common solution is to have nodes broadcast their identity, so that reception at node $A$ of such a beacon from node $B$ suffices for $A$ to add $B$ to its neighbor table. This can be abused by an adversary that forges beacons and misleads a correct, protocol-abiding, node into believing that it has fictitious neighbors. Entity authentication may appear as a solution. Authentication does *not* imply, however, the node is a neighbor. It only establishes which node created a message but not which sent it across the wireless medium. To illustrate this, consider $A$ and $B$ unable to communicate directly, and $C$ within range of both $A$ and $B$. Node $C$ receives and *repeats* $B$'s beacon, for example, digitally signed and time-stamped, with no modification. Then, $A$ receives the beacon and discovers $B$ as a neighbor, even though this is not so. Precisely because $A$ cannot distinguish whether the message (beacon) was sent directly by $B$ or it was *relayed* by another node.

A number of schemes were designed to thwart such relay attacks, often termed *wormholes*, and essentially safeguard ND. *Distance bounding* [2] is the basic approach: the distance of two nodes is estimated by measuring the signal *time of flight* from and to those nodes. If the estimate is below a threshold corresponding to the nodes' communication range, the node is accepted as a neighbor. This may provide the desired level of security for some applications; e.g., if an RFID reader can conclude that a tag is within a range of 10cm, it is safe to have the building door opened. In other words, what this approach provides is discovery of *physical neighborhood*. However, for two nodes to be *communication neighbors* (which we term simply as "neighbors" in the rest of the paper), proximity is *not* sufficient [19]. Obstacles or interference can prevent nearby nodes from communicating directly. This allows the attacker to abuse a ND mechanism oblivious to such obstructions and to mislead two near-by nodes into believing they are neighbors while they are not. This aspect of ND has been largely overlooked by schemes proposed to date.

In this paper, we address this problem, by answering a more fundamental question: To what extent is secure neighbor discovery possible? We focus on the most generally applicable variant of ND, which only requires two nodes to establish a neighbor relation; relying on additional nodes to assist the ND process can be impractical, especially in low-density networks. We prove that for a large class of protocols, which includes many of the proposals in the literature, it is *impossible* to achieve secure ND. On the positive side, we propose a protocol from a different class and prove that it can in fact provide secure neighbor discovery.

To reach this result, we contribute the first formal investigation of secure ND. We provide a model of wireless ad hoc networks rich enough to capture the problem at hand, and a specification of what we term the *two-party* ND. Then, we analyze the above-mentioned two general classes of protocols. We denote the first one *time-based protocols* (*T-protocols*), for which nodes exchange messages and are able to measure time with perfect accuracy. For this class, we show the following *impossibility result*: No T-protocol can solve the (secure) ND problem if adversarial nodes are able to relay messages with a delay below a certain threshold (Section 3). On the contrary, if the minimum relaying delay is above that same threshold, we show it is possible to achieve secure ND (Section 4). Then, in Section 5, we consider the second class of protocols we term *time- and location-based protocols* (*TL-protocols*): nodes are, in addition to T-protocol capabilities, aware of their location. We show that TL-protocols can secure ND even if adversarial nodes can relay messages with almost no delay.

Existing solutions, discussed in Section 7, were not formally analyzed. A fraction of those schemes are indeed affected by our impossibility result. For the rest, our discussion in Section 7 points out other weakness and reflects concepts introduced here. Furthermore, in Section 6, we discuss in detail the implications of our results, model assumptions, as well as practical considerations on protocol design, before we conclude with future work.

## 2. SYSTEM MODEL

We are interested in modeling a wireless network: its basic entities, *nodes*, are processes running on computational platforms equipped with transceivers communicating over a wireless channel. We assume that nodes have synchronized clocks and are static (not mobile). Nodes either follow the implemented system functionality, in which case we denote them as *correct*, or they are under the control of an adversary, in which case we denote them as *adversarial* nodes.

We model communication at the physical layer, rather than at higher layers (data link, network, or application), in order to capture the inherent characteristics of neighbor discovery in wireless networks. For simplicity, correct nodes are assumed to use a single wireless channel and omnidirectional antennas, but we do not require them to have equal transmission power and receiver sensitivity. On the other hand, adversarial nodes have enhanced capabilities: use directional antennas and are able to communicate not only across the wireless channel used by correct nodes, but also across a dedicated *adversary channel* imperceptible to correct nodes.

Our system model comprises: (i) a setting $\mathcal{S}$, which describes the type (correct or adversarial) of nodes, their location and how the wireless channel state changes over time; (ii) a protocol model $\mathcal{P}$, which determines the behavior of correct nodes; (iii) an adversary model $\mathcal{A}$, which determines the capabilities of adversarial nodes.

We make the assumption that if we look at the system at any point in time, one or more phenomena occur. We are interested in phenomena relevant to the wireless communication and the system at hand and, consequently, to our analysis. We denote these phenomena, associated with nodes, as *events* (Definition 2). Then, we model the system evolution over time using the notion of *trace*, i.e., a set of events (Definition 3). More precisely, we use *feasible* traces, that satisfy constraints specified by $\mathcal{S}$ (proper correspondence between wireless sending and receiving of messages), $\mathcal{P}$ (correct nodes follows the protocol), and $\mathcal{A}$ (adversarial nodes behave according to their capabilities).

The specification of secure neighbor discovery is provided exclusively with respect to feasible traces. It consists of two properties requiring that (i) if a node concludes that some other correct node is a neighbor, then it is indeed a neighbor (in every feasible trace), and (ii) if two correct nodes are neighbors, it should be possible for them to conclude they are neighbors (in some setting and feasible trace). We call this *two-party* neighbor discovery, with only two nodes participating in an ND protocol run. We discuss later an alternative *multi-party* ND, which relies on the participation of additional correct nodes to conclude successfully on whether two nodes are neighbors or not.

### 2.1 System Parameters

We list the parameters of our system model. They are used by the protocols, and are known to the protocol designer and to the adversary, both of whom have limited control over their values.

- $\mathbb{V}$, the set of unique *node identifiers*, which for simplicity we will consider equivalent with the nodes themselves,

- $\mathbf{v} \in \mathbb{R}_{>0}$, the *signal propagation speed* across the wireless channel,

- $\mathbf{v}_{\text{adv}} \geqslant \mathbf{v}$, the *information propagation speed* over the *adversary channel*,
- $\mathbb{M}$, the set of *messages*,
- $|.| : \mathbb{M} \to \mathbb{R}_{>0}$, the *message duration* function.

Parameter $\mathbf{v}$ defines how fast messages propagate across the wireless channel, and once a communication technology is selected, this cannot be controlled by the system designer. Parameter $\mathbf{v}_{\text{adv}}$ is under the control of the adversary: he can choose the technology and thus how fast information can propagate between adversarial nodes across the adversary channel. The message space is system-specific and under the control of the system designer, whereas the message duration function, which determines the transmission delay (*not* including the propagation delay), also depends on the technology used and the achievable transmission rates, e.g., in bits per second.

## 2.2 Settings

A setting describes the type and location of nodes, and how the state of the wireless channel changes over time.

DEFINITION 1. *A* setting *$\mathcal{S}$ is a tuple $\langle V, loc, type, link \rangle$, where:*

- *$V \subset \mathbb{V}$ is a finite set of nodes. An ordered pair $(A, B) \in V^2$ is called a* link.
- *$loc : V \to \mathbb{R}^2$ is called a* location *function[1]. As we assume nodes are not mobile, this function does not depend on time. We define $dist : V \times V \to \mathbb{R}_{\geqslant 0}$ as $dist(A, B) = d_2(loc(A), loc(B))$, where $d_2$ is the Euclidean distance in $\mathbb{R}^2$. We require the loc function to be injective, so that no two nodes share the same location. Thus, $dist(A, B) > 0$ for $A \neq B$.*
- *$type : V \to \{correct, adversarial\}$ is the* type *function; it defines which nodes are* correct *and which are* adversarial. *This function does not depend on time, as we assume that the adversary does not corrupt new nodes during the system execution. We denote $V_{\text{cor}} = type^{-1}(\{correct\})$ and $V_{\text{adv}} = type^{-1}(\{adversarial\})$.*
- *$link : V^2 \times \mathbb{R}_{\geqslant 0} \to \{up, down\}$ is the* link state *function. Accordingly to this function we say that at a given time $t \geqslant 0$, a link $(A, B) \in V^2$ is* up *(denoted $t :: A \to B$) or* down *(denoted $t :: A \not\to B$). We use abbreviations $t :: A \leftrightarrow B =_{\text{def}} t :: A \to B \wedge t :: B \to A$ and $t :: A \not\leftrightarrow B =_{\text{def}} t :: A \not\to B \wedge t :: B \not\to A$. We extend the "$t :: A \to B$" notation from single time points to sets as follows: $T :: A \to B =_{\text{def}} \forall t \in T, \ t :: A \to B$. We assume the convention $\mathbb{R}_{\geqslant 0} :: A \not\to A$.*

*We denote the set of all settings by $\Sigma$.*

## 2.3 Traces

We use the notion of *trace* to model an execution of the system. A trace is composed of *events*. We model events related to the wireless communication and the detection of a neighbor. The former, denoted as Bcast, Dcast and Receive, models broadcast (or omnidirectional) transmission, directional transmission, and reception, respectively. The latter,

---
[1] All the results of this paper can be immediately transcribed to $\mathbb{R}^3$. The $\mathbb{R}^2$ space is used only for presentation simplicity.

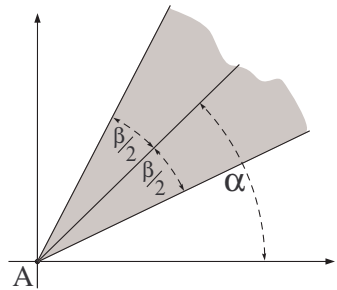

**Figure 1: Range of the Dcast primitive. $inrange(A, \alpha, \beta, B)$ is true iff $B$ is located in the gray region.**

denoted as Neighbor, means that a node accepts another node as a neighbor. Each event is primarily associated with (essentially, takes place at) a node we denote as the *active* node. For some events, a secondary association with another node can exist. In particular:

DEFINITION 2. *An* event *is one of the following terms:*

- $\text{Bcast}(A; t; m)$
- $\text{Dcast}(A; t; \alpha, \beta, m)$
- $\text{Receive}(A; t; B, m)$
- $\text{Neighbor}(A; t; C, t')$

*where: $A \in \mathbb{V}$ is the* active node, *$t \in \mathbb{R}_{\geqslant 0}$ is the* start time, *$m \in \mathbb{M}$ is a* message, *$\alpha \in [0, 2\pi)$ is the* sending direction, *$\beta \in (0, 2\pi]$ is the* sending angle, *$B \in \mathbb{V}$ is the* sender node, *$C \in \mathbb{V}$ is a* declared neighbor, *$t' \in \mathbb{R}_{\geqslant 0}$ is the time at which $C$ is a neighbor according to $A$'s declaration.*

*For an event $e$, we write $start(e)$ for its start time and $end(e)$ for its end time. For events including a message $m$, $end(e) = start(e) + |m|$, while for the Neighbor event $end(e) = start(e)$.*

Dcast, representing a message sent with a directional antenna at direction $\alpha$ over an angle $\beta$, is illustrated in Figure 1. Receive represents message reception caused (triggered) by any incoming message, and thus a previous Bcast and Dcast event (self-triggered). Neighbor can be thought of as an internal outcome of a neighbor discovery protocol (to be defined later). Then, traces comprising the above events are defined.

DEFINITION 3. *A* trace *$\theta$ is a set of events that satisfies what we will call the* finite cut condition: *for any finite $t \geqslant 0$, the subset $\{e \in \theta \mid start(e) < t\}$ is finite.*

*We denote the set of all traces by $\Theta$.*

The finite cut condition ensures that during any finite interval of time only a finite number of events occurs; as settings comprise a finite number of nodes, this is natural to demand.

## 2.4 Setting-Feasible Traces

Feasibility with respect to a setting $\mathcal{S}$ is a set of conditions ensuring a proper causal and time relation between send and receive events.

DEFINITION 4. *A trace $\theta \in \Theta$ is* feasible *with respect to a setting $\mathcal{S} = \langle V, loc, type, link \rangle$, if the following conditions are satisfied:*

1. $\forall \mathsf{Receive}(A; t; B, m) \in \theta$,
   $(A, B \in V) \wedge ([t, t + |m|] :: B {\rightarrow} A) \wedge$
   $(\mathsf{Bcast}(B; t - t_{AB}; m) \in \theta \veebar (inrange(B, \alpha, \beta, A) \wedge$
   $\mathsf{Dcast}(B; t - t_{AB}; \alpha, \beta, m) \in \theta))$

2. $\forall \mathsf{Bcast}(A; t; m) \in \theta$, $(A \in V) \wedge$
   $(\forall B \in V, \ [t + t_{AB}, t + t_{AB} + |m|] :: A {\rightarrow} B \implies$
   $\mathsf{Receive}(B; t + t_{AB}; A, m) \in \theta)$

3. $\forall \mathsf{Dcast}(A; t; \alpha, \beta, m) \in \theta$, $(A \in V_{\mathrm{adv}}) \wedge$
   $(\forall B \in V, \ (inrange(A, \alpha, \beta, B) \wedge$
   $[t + t_{AB}, t + t_{AB} + |m|] :: A {\rightarrow} B) \implies$
   $\mathsf{Receive}(B; t + t_{AB}; A, m) \in \theta)$

*Where $\veebar$ denotes logical exclusive or, $t_{AB} = \frac{dist(A,B)}{\mathbf{v}}$ is the time of flight, and $inrange(A, \alpha, \beta, B)$ is defined in Figure 1.*

*We denote the set of all traces feasible with respect to a setting $\mathcal{S}$ by $\Theta_{\mathcal{S}}$.*

Condition 1 of Definition 4 ensures that every message that is received was previously sent. Condition 2 ensures that a broadcasted message is received by all nodes enabled to do so by the link relation.[2] Condition 3 ensures that a Dcast-ed message is received only by the nodes in the area as per the Dcast transmission (see Figure 1) and only if the link is up. In other words, communication is causal (a receive is always preceded by a sent), and reliable *as long as the link is up*. Unreliability, expected and common in wireless communications, is modeled by the state of the link being *down*. Furthermore, the three conditions in Definition 4 introduce a strict time relation between events, reflecting line-of-sight signal propagation across the channel with a constant speed $\mathbf{v}$.

## 2.5 Protocol-Feasible Traces

A trace is essentially a *global view* of the system execution. To describe what a node observes during a system execution, we use the notion of *local view*, primarily comprising a *local trace* composed of *local events*. We define these next.

DEFINITION 5. *A* local event *is one of the terms:* $\mathsf{Bcast}(t; m)$, $\mathsf{Receive}(t; m)$, $\mathsf{Neighbor}(t; B, t')$, *where $B \in \mathbb{V}$, $m \in \mathbb{M}$, $t, t' \in \mathbb{R}_{\geqslant 0}$. For a local event $e$, $start(e)$, $end(e)$ are defined as in Definition 2.*

DEFINITION 6. *A* local trace *is a set of local events that satisfies the finite cut condition. Given a node identifier $A \in \mathbb{V}$, time $t \geqslant 0$ and trace $\theta \in \Theta$, we calculate the* local trace *of node $A$ at time $t$ in trace $\theta$, denoted $\theta|_{A,t}$, as follows:*

$$\theta|_{A,t} = \{\mathsf{Bcast}(t_1; m) \mid t_1 < t \wedge$$
$$\mathsf{Bcast}(A; t_1; m) \in \theta\} \cup \quad (1)$$
$$\{\mathsf{Receive}(t_1; m) \mid t_1 + |m| < t \wedge$$
$$\exists B \in \mathbb{V}, \ \mathsf{Receive}(A; t_1; B, m) \in \theta\} \cup \quad (2)$$
$$\{\mathsf{Neighbor}(t_1; B, t') \mid t_1 < t \wedge$$
$$\mathsf{Neighbor}(A; t_1; B, t') \in \theta\} \quad (3)$$

---
[2] Note that time is "measured" at the receiver, not the sender.

*We call $\theta|_{A,\infty}$ a* complete local trace *of $A$ in $\theta$ and denote it shortly $\theta|_A$.*

Note that the Receive local event, contrary to its global counterpart, does not include the information about the sender of the message. This is of central importance, capturing the earlier mentioned fundamental challenge in securing ND in wireless networks: the receiver of a message cannot reliably identify who the sender is. This is because identifiers included in a message can be forged, and even cryptography can at most allow to identify the creator of a message, not the sender.

We identify two variants of the local view notion: an *T-local view*, as the basis for defining the class of time-based protocols, and an *TL-local view*, used to define the class of time- and location-based protocols.

DEFINITION 7. *Given a trace $\theta$, an* T-local view *of node $A$ at time $t$ in $\theta$ is a tuple $\langle A, t, \theta|_{A,t} \rangle$; we denote it $\theta||_{A,t}$.*

DEFINITION 8. *Given a trace $\theta$ and a setting $\mathcal{S}$, an* TL-local view *of node $A$ at time $t$ in trace $\theta$ is a tuple $\langle A, t, loc(A), \theta|_{A,t} \rangle$; we denote it $\theta||_{\mathcal{S},A,t}$, or $\theta||_{A,t}$ is setting $\mathcal{S}$ is clear from the context.*

Note that $\mathcal{S}$ is part of Definition 8 as the location of node $A$ is defined only within a specific setting. With the notion of local view(s) in hand, we can proceed with the definition of a protocol model. This definition captures the property of protocols essential to our investigation: the fact that protocol behavior depends *exclusively* on the local view of the node executing the protocol.

DEFINITION 9. *An* T(TL)-protocol model $\mathcal{P}$ *is a function which given a T(TL)-local view $\theta||_{A,t}$, determines a finite, non-empty set of* actions*; an* action *is one of the terms:* $\epsilon$, $\mathsf{Bcast}(m)$ *or* $\mathsf{Neighbor}(A, t)$, *where $m \in \mathbb{M}, A \in \mathbb{V}, t \in \mathbb{R}_{\geqslant 0}$.*

The interpretation of Bcast and Neighbor actions is natural. The $\epsilon$ action means that the node does not execute an event, with the exception of possible Receive event(s). Note that modeling the protocol output (i.e., the protocol model codomain) as a family of *sets of actions* allows for non-deterministic protocols.

DEFINITION 10. *A trace $\theta \in \Theta_{\mathcal{S}}$ is* feasible *with respect to a T- or TL-protocol model $\mathcal{P}$, if the following conditions are satisfied:*

1. $\forall A \in V_{\mathrm{cor}}, \ \forall \mathsf{Bcast}(A; t; m) \in \theta, \ \mathsf{Bcast}(m) \in \mathcal{P}(\theta||_{A,t})$

2. $\forall A \in V_{\mathrm{cor}}, \ \forall \mathsf{Neighbor}(A; t; B, t') \in \theta$,
   $\mathsf{Neighbor}(B, t') \in \mathcal{P}(\theta||_{A,t})$

3. $\forall A \in V_{\mathrm{cor}}, \ \forall t \in X_A, \ \epsilon \in \mathcal{P}(\theta||_{A,t})$, where
   $X_A = \mathbb{R}_{\geqslant 0} \setminus start(\theta|_A \cap E)$,
   $E = \{\mathsf{Bcast}(t; m) \mid m \in \mathbb{M}, t \in \mathbb{R}_{\geqslant 0}\} \cup$
   $\{\mathsf{Neighbor}(t; B, t') \mid B \in \mathbb{V}, t, t' \in \mathbb{R}_{\geqslant 0}\}$

*We denote the set of all traces feasible with respect to a setting $\mathcal{S}$ and T(TL)-protocol model $\mathcal{P}$ by $\Theta_{\mathcal{S},\mathcal{P}}$.*

Conditions 1 and 2 of Definition 10 ensure that Bcast of Neighbor actions taken by a node are allowed by the protocol. Condition 3, with $X_A$ the set of all points in time at which no event other than Receive happens at node $A$, ensures that the protocol allows for a node to not perform an action.

## 2.6 Adversary-Feasible Traces

For the purpose of the impossibility result, we consider first a relatively limited adversary, that is only capable of relaying messages. We denote this model as $\mathcal{A}_{\Delta_{\text{relay}}}$, with the $\Delta_{\text{relay}} > 0$ parameter the minimum relaying delay introduced by an adversarial node; this delay is due to processing exclusively, it does not include any propagation time.

DEFINITION 11. *A trace $\theta \in \Theta_{\mathcal{S},\mathcal{P}}$ is feasible with respect to an adversary model $\mathcal{A}_{\Delta_{\text{relay}}}$ if:*

1. $\forall \mathsf{Bcast}(A;t;m) \in \theta, \ A \notin V_{\text{adv}}$

2. $\forall \mathsf{Dcast}(A;t;\alpha,\beta,m) \in \theta, \ \exists B \in V_{\text{adv}}$,
   $\exists \delta \geqslant \Delta_{\text{relay}} + \frac{dist(A,B)}{\mathbf{v}_{\text{adv}}}, \ \exists C \in V$,
   $\mathsf{Receive}(B; t - \delta; C, m) \in \theta$

*We denote the set of all traces feasible with respect to a setting $\mathcal{S}$, T-protocol model $\mathcal{P}$, and adversary model $\mathcal{A}_{\Delta_{\text{relay}}}$ by $\Theta_{\mathcal{S},\mathcal{P},\mathcal{A}_{\Delta_{\text{relay}}}}$.*

Condition 1 of Definition 11 is only to facilitate the presentation of proofs in subsequent sections, stating that adversarial nodes do not use the Bcast primitive. Condition 2 states that every message sent by an adversarial node is necessarily a replay of a message $m$ that either this or another adversarial node received. In addition, the delay between receiving $m$ and re-sending it, or more precisely the difference between the start times of the corresponding events, needs to be at least $\Delta_{\text{relay}}$, plus the propagation delay across the adversary channel in case another adversarial node received the relayed message.

From $\mathcal{A}_{\Delta_{\text{relay}}}$, we derive two weaker adversary models, $\mathcal{A}'_{\Delta_{\text{relay}}}$ and $\mathcal{A}''_{\Delta_{\text{relay}}}$, defined next. Model $\mathcal{A}'_{\Delta_{\text{relay}}}$ restricts adversarial nodes to broadcasts, while $\mathcal{A}''_{\Delta_{\text{relay}}}$ precludes adversarial nodes from utilizing an adversary channel. As it will become clear in Section 3, all these adversary models are valuable for the impossibility result, and their weakness *strengthens* the impossibility result.

DEFINITION 12. *A trace $\theta \in \Theta_{\mathcal{S},\mathcal{P}}$ is feasible with respect to an adversary model $\mathcal{A}'_{\Delta_{\text{relay}}}$ if:*

1. $\forall \mathsf{Dcast}(A;t;\alpha,\beta,m) \in \theta, \ A \notin V_{\text{adv}}$

2. $\forall \mathsf{Bcast}(A;t;m) \in \theta \in \theta, \ \exists B \in V_{\text{adv}}$,
   $\exists \delta \geqslant \Delta_{\text{relay}} + \frac{dist(A,B)}{\mathbf{v}_{\text{adv}}}, \ \exists C \in V$,
   $\mathsf{Receive}(B; t - \delta; C, m) \in \theta$

DEFINITION 13. *A trace $\theta \in \Theta_{\mathcal{S},\mathcal{P}}$ is feasible with respect to an adversary model $\mathcal{A}''_{\Delta_{\text{relay}}}$ if:*

1. $\forall \mathsf{Bcast}(A;t;m) \in \theta, \ A \notin V_{\text{adv}}$

2. $\forall \mathsf{Dcast}(A;t;\alpha,\beta,m) \in \theta, \ \exists \delta \geqslant \Delta_{\text{relay}}$,
   $\exists C \in V, \ \mathsf{Receive}(A; t - \delta; C, m) \in \theta$

## 2.7 Neighbor Discovery Specification

The ability to communicate directly, without the intervention or 'assistance' of relays, is expressed in our model by a link being up, thus the following definition:

DEFINITION 14. *Node $A$ is a neighbor of node $B$ in setting $\mathcal{S}$ at time $t$, if $t::A \rightarrow B$. If $t::A \leftrightarrow B$ we will say that nodes $A$ and $B$ are neighbors at time $t$.*

For simplicity of presentation, we use the "$t::A \rightarrow B$" notation to denote the neighbor relation, as well as the link relation. Having defined the neighbor relation, we are ready to present the formal specification of secure neighbor discovery. This definition uses a parameter: $\mathbf{R} \in \mathbb{R}_{>0}$, the *neighbor discovery (ND) range*. Typically, $\mathbf{R}$ is equal to the *nominal communication range* for a given wireless medium, however, we use $\mathbf{R}$ more freely as the communication range for which ND inferences are drawn.

DEFINITION 15. *A protocol model $\mathcal{P}$ satisfies(solves) two-party neighbor discovery for an adversary model $\mathcal{A}$, if the following properties are both satisfied:*

ND1 $\forall \mathcal{S} \in \Sigma, \ \forall \theta \in \Theta_{\mathcal{S},\mathcal{P},\mathcal{A}}, \ \forall A, B \in V_{\text{cor}}$,
$\mathsf{Neighbor}(A;t;B,t') \in \theta \implies t'::B \rightarrow A$

ND2 $\forall d \in (0, \mathbf{R}], \ \forall A, B \in \mathbb{V}, A \neq B, \ \exists \mathcal{S} \in \Sigma$,
$V = V_{\text{cor}} = \{A, B\} \ \wedge \ dist(A,B) = d \ \wedge \ \mathbb{R}_{\geqslant 0}::A \leftrightarrow B$
$\wedge \ \exists \theta \in \Theta_{\mathcal{S},\mathcal{P},\mathcal{A}}, \ \mathsf{Neighbor}(A;t;B,t') \in \theta$

Intuitively, property ND1 requires that if a node accepts some other correct node $B$ as a neighbor at time $t'$, then $B$ is actually a neighbor at that time. Property ND2 complements ND1, assuring that the protocol offers minimal *availability*: it requires that for every distance $d$ in the desired ND range $\mathbf{R}$, there should be at least some setting, in which the protocol is able to conclude that a node is a neighbor (in some, not all executions); this setting should contain exactly two nodes at distance $d$, being neighbors, and both *correct*. The "two-nodes setting" requirement clarifies why we call this *two-party* ND. The ND2 property is the least that can be required from a usable two-party ND protocol: indeed, a protocol not satisfying this property would be unable to conclude, for some distance(s) in the ND range, that nodes are neighbors. This makes the impossibility result in Section 3 more meaningful: impossibility with respect to a weak property implies impossibility for any stronger property.

## 3. IMPOSSIBILITY FOR T-PROTOCOLS

We show in this section that no time-based protocol can solve the two-party neighbor discovery problem as specified by Definition 15. We base the proof on the fact, captured in Lemma 1, that it is impossible for a correct node to distinguish between different settings based on an T-local view. The impossibility result in Theorem 1 stems from showing two settings which are indistinguishable by a correct node, one in which two nodes are neighbors and one where they are not. We elaborate on the assumptions and implications of this result in Section 6.

We emphasize that the non-restricted form of the message space $\mathbb{M}$ encompasses all possible messages including, for example, time-stamps and any type of cryptography, thus contributing to the generality of the impossibility result.

LEMMA 1. *Let $\mathcal{P}$ be a T-protocol model, $\mathcal{S}$ and $\mathcal{S}'$ be settings such that $V_{\text{cor}} = V'_{\text{cor}}$, and $\theta \in \Theta_{\mathcal{S},\mathcal{P}}$ and $\theta' \in \Theta_{\mathcal{S}'}$ be traces such that local traces $\theta|_A = \theta'|_A$ for all $A \in V_{\text{cor}}$. Then $\theta'$ is feasible with respect to T-protocol model $\mathcal{P}$.*

The proof of Lemma 1 can be found in [22].

THEOREM 1. *There exists no T-protocol model that satisfies two-party neighbor discovery for the adversary model $\mathcal{A}''_{\Delta_{\text{relay}}}$ if $\Delta_{\text{relay}} < \frac{\mathbf{R}}{\mathbf{v}}$.*

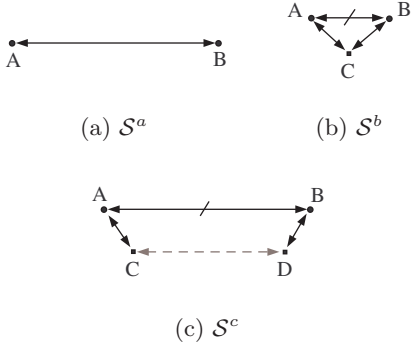

**Figure 2: Settings used in the impossibility result proof.** Settings $\mathcal{S}^a = \langle \{A,B\}, loc^a, type^a, link^a \rangle$, $\mathcal{S}^b = \langle \{A,B,C\}, loc^b, type^b, link^b \rangle$ and $\mathcal{S}^c = \langle \{A,B,C,D\}, loc^c, type^c, link^c \rangle$. In all settings, nodes $A$ and $B$ are correct, nodes $C$ and $D$ are adversarial. The location functions are such that $dist^b(A,C) + dist^b(B,C) + \mathbf{v}\Delta_{\text{relay}} \leqslant dist^a(A,B) \leqslant \mathbf{R}$ and $dist^c(A,C) + dist^c(D,B) + \frac{\mathbf{v}}{\mathbf{v}_{\text{adv}}} dist^c(C,D) + \mathbf{v}\Delta_{\text{relay}} \leqslant dist^a(A,B)$. The state of links does not change over time and is shown in the figure. The dashed arrow in figure (c) denotes the adversarial channel.

PROOF. To prove that under the assumptions of the theorem no T-protocol model can satisfy both ND1 and ND2, we show that any T-protocol model that satisfies ND2 cannot satisfy ND1.

Take any T-protocol model $\mathcal{P}$ satisfying ND2. Pick some distance $\geqslant \mathbf{v}\Delta_{\text{relay}}$ in the ND range. Property ND2 guarantees the existence of a setting such as the one shown in Figure 2(a) (we denote it $\mathcal{S}^a$) and the existance of a trace $\theta \in \Theta_{\mathcal{S}^a, \mathcal{P}, \mathcal{A}_{\Delta_{\text{relay}}}}$ such that $\mathsf{Neighbor}(A; t; B, t') \in \theta$. As $\theta$ is feasible with respect to setting $\mathcal{S}^a$, this trace has to be of the form:

$$\theta = \{\mathsf{Bcast}(A; t_i; m_i) \mid i \in I_A\} \cup$$
$$\{\mathsf{Receive}(B; t_i + \Delta; A, m_i) \mid i \in I_A\} \cup$$
$$\{\mathsf{Bcast}(B; t_i; m_i) \mid i \in I_B\} \cup$$
$$\{\mathsf{Receive}(A; t_i + \Delta; B, m_i) \mid i \in I_B\} \cup$$
$$\{\mathsf{Neighbor}(A; t_i; B, t'_i) \mid i \in J_A\} \cup$$
$$\{\mathsf{Neighbor}(B; t_i; A, t'_i) \mid i \in J_B\}$$

where $\Delta = \frac{dist^a(A,B)}{\mathbf{v}}$, $t_i, t'_i \in \mathbb{R}_{\geqslant 0}$ and $I_A, I_B, J_A, J_B$ are pairwise disjoint index sets with $J_A \neq \emptyset$ (all the other index sets can be empty).

In setting $\mathcal{S}^b$, shown in figure 2(b), we have $\mathbb{R}_{\geqslant 0} :: B \leftrightarrow A$. Consider the following trace $\theta'$, which is is essentially the same as $\theta$, but for node $C$ relaying all the communication between nodes $A$ and $B$:

$$\theta' = \{\mathsf{Bcast}(A; t_i; m_i) \mid i \in I_A\} \cup$$
$$\{\mathsf{Receive}(C; t_i + \delta_1; A, m_i) \mid i \in I_A\} \cup$$
$$\{\mathsf{Dcast}(C; t_i + \delta_2; 0, \pi, m_i) \mid i \in I_A\} \cup$$
$$\{\mathsf{Receive}(B; t_i + \Delta; C, m_i) \mid i \in I_A\} \cup$$
$$\{\mathsf{Bcast}(B; t_i; m_i) \mid i \in I_B\} \cup$$
$$\{\mathsf{Receive}(C; t_i + \delta_3; B, m_i) \mid i \in I_B\} \cup$$
$$\{\mathsf{Dcast}(C; t_i + \delta_4; -\pi, \pi, m_i) \mid i \in I_B\} \cup$$
$$\{\mathsf{Receive}(A; t_i + \Delta; C, m_i) \mid i \in I_B\} \cup$$
$$\{\mathsf{Neighbor}(A; t_i; B, t'_i) \mid i \in J_A\} \cup$$
$$\{\mathsf{Neighbor}(B; t_i; A, t'_i) \mid i \in J_B\}$$

where $\delta_1 = \frac{dist^b(A,C)}{\mathbf{v}}$, $\delta_2 = \Delta - \frac{dist^b(C,B)}{\mathbf{v}}$, $\delta_3 = \frac{dist^b(B,C)}{\mathbf{v}}$ and $\delta_4 = \Delta - \frac{dist^b(C,A)}{\mathbf{v}}$.

It is simple to check that this trace is feasible with respect to setting $\mathcal{S}^b$. It is also feasible with respect to T-protocol model $\mathcal{P}$: as $\theta|_{A,t} = \theta'|_{A,t}$ and $\theta|_{B,t} = \theta'|_{B,t}$, this follows from Lemma 1. Finally, $\theta'$ is feasible with respect to the adversary model $\mathcal{A}_{\Delta_{\text{relay}}}$, because $\delta_2 - \delta_1 = \delta_4 - \delta_3 \geqslant \Delta_{\text{relay}}$. Therefore $\theta'$ belongs to $\Theta_{\mathcal{S}^b, \mathcal{P}, \mathcal{A}_{\Delta_{\text{relay}}}}$ and together with $\mathcal{S}^b$ forms the counterexample that we were looking for: $A$ concludes $B$ is a neighbor whereas it is not. Thus, T-protocol model $\mathcal{P}$ does not satisfy ND1. As $\mathcal{P}$ was chosen arbitrarily, this concludes the proof. $\square$

We can use the same technique (using settings $\mathcal{S}^a$ and $\mathcal{S}^c$, illustrated in Figure 2) to prove a corresponding theorem for the adversary model $\mathcal{A}'_{\Delta_{\text{relay}}}$:

THEOREM 2. *There exists no T-protocol model that satisfies two-party neighbor discovery for the adversary model $\mathcal{A}'_{\Delta_{\text{relay}}}$ if $\Delta_{\text{relay}} < \frac{\mathbf{R}}{\mathbf{v}}$.*

## 4. T-PROTOCOL SOLVING ND

Theorem 1 considers adversarial nodes that relay messages with a delay smaller than $\frac{\mathbf{R}}{\mathbf{v}}$. In this section we demonstrate a specific T-protocol (we denote it as $\mathcal{P}^\mathsf{T}$), which satisfies ND (Definition 15) if the minimum relaying delay incurred by adversarial nodes is greater than $\frac{\mathbf{R}}{\mathbf{v}}$ (Theorem 3, the proof can be found in [22]).

*Protocol.*

Informally, the $\mathcal{P}^\mathsf{T}$ protocol requires nodes to transmit authenticated messages containing a time-stamp set at the time of sending. Upon receipt of such a message, a receiver checks its "freshness" by verifying that the message time-stamp is within a threshold of the receiver's current time. If so, it accepts the message creator as a neighbor. Note that this protocol is essentially the *temporal packet leash* proposed by Hu, Perrig and Johnson in [13].

*Message Space.*

We specify the message space relevant to this particular T-protocol to be:

$$\{\mathsf{auth}_A(t)\}_{A \in \mathbb{V}, t \in \mathbb{R}_{\geqslant 0}} \subseteq \mathbb{M}$$

with $\mathsf{auth}_A(x)$ denoting that the content of message $x$ is authenticated by node $A$. We do not dwell on which cryp-tographic primitive (e.g., digital signature or message authentication code) is used to this end. We call the message $\mathsf{auth}_A(t)$ a *beacon message*, and $t$ the *beacon-time*.

*Feasibility.*

Below we define feasibility with respect to protocol $\mathcal{P}^\mathsf{T}$ described informally above.[3]

DEFINITION 16. *A trace $\theta \in \Theta_\mathcal{S}$ is feasible with respect to $\mathcal{P}^\mathsf{T}$, if the following conditions are satisfied:*

1. $\forall A \in V_{\text{cor}}, \quad \forall \mathsf{Bcast}(A; t_1; \mathsf{auth}_B(t)) \in \theta,$
   $(B = A) \wedge (t = t_1)$

2. $\forall A \in V_{\text{cor}}, \quad \forall \mathsf{Neighbor}(A; t_0; B, t_1) \in \theta, \quad \exists C \in V,$
   $(\mathsf{Receive}(A; t_1; C, \mathsf{auth}_B(t)) \in \theta) \wedge (t_1 - t \leqslant \frac{\mathbf{R}}{\mathbf{v}}) \wedge$
   $(t_0 > end(\mathsf{Receive}(A; t_1; C, \mathsf{auth}_B(t))))$

Condition 1 ensures that a correct node only broadcasts beacon messages that are authenticated by itself and that have the beacon-time set to the start of the beacon sending time. Recall that correct nodes have synchronized clocks, otherwise they cannot be considered correct. Condition 2 ensures that a correct $A$ accepts $B$ as a neighbor only after it receives and deems fresh a beacon generated by $B$.

*Adversary Model.*

Towards proving that $\mathcal{P}^\mathsf{T}$ solves the ND problem, we need to develop a stronger than $\mathcal{A}_{\Delta_{\text{relay}}}$ adversary model. This is necessary, as proving that a protocol is secure against a weak adversary would be of little value. The new adversary model, $\mathcal{A}^\mathsf{T}_{\Delta_{\text{relay}}}$, allows for not only message relay but also for generation and transmission of any message, as long as the employed cryptosystem is not broken (this approach is compliant with the classical Dolev-Yao model [6]).

DEFINITION 17. *A trace $\theta \in \Theta_{\mathcal{S}, \mathcal{P}^\mathsf{T}}$ is feasible with respect to an adversary model $\mathcal{A}^\mathsf{T}_{\Delta_{\text{relay}}}$ if:*

1. $\forall \mathsf{Bcast}(A; t; m) \in \theta, \quad A \notin V_{\text{adv}}$

2. $\forall A \in V_{\text{adv}}, \quad \forall \mathsf{Dcast}(A; t_1; \alpha, \beta, \mathsf{auth}_B(t)) \in \theta,$
   $(B \in V_{\text{adv}}) \vee (\exists C \in V_{\text{adv}}, \quad \exists \delta \geqslant \Delta_{\text{relay}} + \frac{dist(C,A)}{\mathbf{v}_{\text{adv}}},$
   $\exists D \in V, \quad \mathsf{Receive}(C; t_1 - \delta; D, \mathsf{auth}_B(t)) \in \theta)$

Condition 1 simplifies the presentation mandating that adversarial nodes do not use the $\mathsf{Bcast}$ primitive. Nonetheless, this is not a limitation because $\mathsf{Bcast}(m)$ is equivalent to $\mathsf{Dcast}(0, 2\pi, m)$, by which we mean that it triggers exactly the same $\mathsf{Receive}(m)$ events. Condition 2 ensures that an adversarial node is allowed to send any message as long as it is authenticated by an adversarial node (itself or other). This implies that adversarial nodes can share cryptographic keys or any material used for authentication. Furthermore, Condition 2 reflects that the adversary cannot forge authenticated messages: it ensures that a message sent by an adversarial node, and authenticated by a correct node must be a relayed one. In other words, some (possibly the same) adversarial node must have received this message earlier, at least $\Delta_{\text{relay}}$ plus the propagation time between the two nodes (over the adversarial channel).

THEOREM 3. *If $\Delta_{\text{relay}} \geqslant \frac{\mathbf{R}}{\mathbf{v}}$ then $\mathcal{P}^\mathsf{T}$ satisfies neighbor discovery for the adversary model $\mathcal{A}^\mathsf{T}_{\Delta_{\text{relay}}}$.*

---

[3]For clarity and brevity, we define this "from scratch," rather than specifying an T-protocol model according to Definition 9 and relying on Definition 10 for feasibility.

## 5. TL-PROTOCOL SOLVING ND

Time- and location-based protocols, compared to the T-protocol class, augment nodes with location awareness. Because nodes are more powerful, we can show that if $\mathbf{v} = \mathbf{v}_{\text{adv}}$, an TL-protocol we denote as $\mathcal{P}^\mathsf{GT}$ solves ND regardless of how small $\Delta_{\text{relay}}$ is. The reason the impossibility theorem does not apply can be traced back to Lemma 1: even given identical local traces, correct nodes can resort to location information to distinguish setting $\mathcal{S}^a$ from $\mathcal{S}^b$. The proof is similar to that of the T-protocol case, found in [22].

*Protocol.*

Informally, the $\mathcal{P}^\mathsf{GT}$ protocol requires that nodes send authenticated messages containing a time-stamp set at the time of sending and their own location. Upon receipt of such a message $m$ sent from a node $B$, the receiver $A$ calculates two estimates of the $A, B$ distance. The first estimate is based on the difference of its own clock at reception time (the start of reception) and $m$'s time-stamp. The second one is calculated with the help of the location in $m$ and $A$'s location. If the two distance estimates are equal, and $m$ is authenticated, $A$ accepts $B$ as a neighbor. Note that this protocol is a combination between the *temporal* and the *geographical packet leash* [13].

*Message Space.*

We specify the message space as follows:

$$\{\mathsf{auth}_A(t, l)\}_{A \in \mathbb{V}, t \in \mathbb{R}_{\geqslant 0}, l \in \mathbb{R}^2} \subseteq \mathbb{M}$$

We call the message $\mathsf{auth}_A(t, l)$ a *beacon message*, $t$ the *beacon-time* of the message, and $l$ the *beacon-location* of the message.

*Feasibility.*

The following defines feasibility with respect to $\mathcal{P}^\mathsf{GT}$.

DEFINITION 18. *A trace $\theta \in \Theta_\mathcal{S}$ is feasible with respect to $\mathcal{P}^\mathsf{GT}$, if the following conditions are satisfied:*

1. $\forall A \in V_{\text{cor}}, \quad \forall \mathsf{Bcast}(A; t_1; \mathsf{auth}_B(t, l)) \in \theta,$
   $B = A \wedge t = t_1 \wedge l = loc(A)$

2. $\forall A \in V_{\text{cor}}, \quad \forall \mathsf{Neighbor}(A; t_0; B, t_1) \in \theta, \quad \exists C \in V,$
   $\mathsf{Receive}(A; t_1; C, \mathsf{auth}_B(t, l)) \in \theta \wedge t_1 - t = \frac{d_2(loc(A), l)}{\mathbf{v}}$
   $\wedge \ t_0 > end(\mathsf{Receive}(A; t_1; C, \mathsf{auth}_B(t, l)))$

*Adversary Model.*

The adversary model, denoted $\mathcal{A}^\mathsf{GT}_{\Delta_{\text{relay}}}$, is almost identical to $\mathcal{A}^\mathsf{T}_{\Delta_{\text{relay}}}$ but for the format of beacon messages.

DEFINITION 19. *A trace $\theta \in \Theta_{\mathcal{S}, \mathcal{P}^\mathsf{GT}}$ is feasible with respect to the adversary model $\mathcal{A}^\mathsf{GT}_{\Delta_{\text{relay}}}$ if:*

1. $\forall \mathsf{Bcast}(A; t; m) \in \theta, \quad A \notin V_{\text{adv}}$

2. $\forall A \in V_{\text{adv}}, \quad \forall \mathsf{Dcast}(A; t_1; \alpha, \beta, \mathsf{auth}_B(t, l)) \in \theta,$
   $(B \in V_{\text{adv}}) \vee (\exists C \in V_{\text{adv}}, \quad \exists \delta \geqslant \Delta_{\text{relay}} + \frac{dist(C,A)}{\mathbf{v}_{\text{adv}}},$
   $\exists D \in V, \quad \mathsf{Receive}(C; t_1 - \delta; D, \mathsf{auth}_B(t, l)) \in \theta)$

THEOREM 4. *If $\mathbf{v} = \mathbf{v}_{\text{adv}}$ and $\Delta_{\text{relay}} > 0$ then $\mathcal{P}^\mathsf{GT}$ satisfies neighbor discovery for the adversary model $\mathcal{A}^\mathsf{GT}_{\Delta_{\text{relay}}}$.*

# 6. DISCUSSION

## 6.1 Implications

The impossibility result points to a fundamental limitation in securing communication ND with T-protocols. Any T-protocol, regardless of the node clock accuracy or processing power, can be attacked by an adversary capable of relaying messages with a small enough delay. As we discuss in the next paragraph, the space for attacks can seem relatively small if $\mathbf{v} = \mathbf{v}_{\text{adv}}$. Nevertheless, it *can* be large enough to constitute a realistic threat, depending essentially on three factors. One of these is very specific to the operational environment, and deals with the following question: How probable is it to have no link between two nodes at distance $d$? This is because for every non-existing link the adversary can set up a short-range relay attack.

For the two other factors, we turn to theorems 1 and 3. These show that for an attack to be successful, the relaying delay of the adversary has to be below the threshold $\frac{\mathbf{R}}{\mathbf{v}}$. This implies the second factor - the expected threat level. If the system designer aims at protecting the network only against relatively limited, slow-relaying adversaries, T-protocols can provide sufficient security (details in Section 6.3). The third factor is the ND range $\mathbf{R}$. In some cases, the system designer might be able to select a low $\mathbf{R}$: this forces the adversary to relay messages faster, but it also precludes the discovery of nodes that are directly reachable but farther than $\mathbf{R}$. Nonetheless, $\mathbf{R}$ needs to be typically equal to the communication range. Thus, for some wireless technologies, ND using T-protocols will be more vulnerable than for others. For example, if we can consider relatively short-range 802.11 radios, communicating typically at 100 to 150$m$, the threshold is $\frac{100m}{c} \approx 333ns$, still significantly above the feasible 40$ns$ relaying delay reported by [24]. For WiMAX, with a range up to 50$km$, the threshold is around 166$\mu s$ leaving much more space for attacks. In fact, as $\mathbf{R} \to \infty$, T-protocols become useless for securing ND, if obstacles can be present in the environment.

In short, T-protocols need to be used with a lot of caution to secure ND. Unless there are no obstacles in the environment, the ND range is low, or only slow-relaying adversaries are of concern, T-protocols cannot provide reliable security, as they are able to prevent only wormholes ranging beyond $\mathbf{R}$. For generally applicable secure ND it is necessary to go beyond the T-protocol class. As Theorem 4 shows, one possibility is the TL-class with protocols such as $\mathcal{P}^{\text{GT}}$ which can secure ND regardless of $\Delta_{\text{relay}}$ or $\mathbf{R}$. Unfortunately, $\mathcal{P}^{\text{GT}}$ is more demanding on the nodes (location awareness), and it requires line-of-sight communication (Section 6.3).

*Simple Quantitative Results.*

Theorem 1 and Theorem 2 show that it is impossible to secure ND even if the adversary cannot utilize an adversarial channel for the communication of the nodes it controls (but in that case it uses directional antennas). However, quantitatively, the relative magnitude of $\mathbf{v}$ and $\mathbf{v}_{\text{adv}}$, the signal propagation velocity across the system wireless channel and the adversary channel, respectively, determine the impact of the adversary.

To illustrate this, we consider first the $\mathcal{A}''_{\Delta_{\text{relay}}}$ adversary and the $\mathcal{S}^b$ setting in Figure 2, with $A, B$ correct and $C$ adversarial nodes, for which $dist^b(A,C) + dist^b(B,C) + \mathbf{v}\Delta_{\text{relay}} \leqslant \mathbf{R}$. These conditions are necessary for the attack to be possible. The last inequality yields, when combined with the triangle inequality $dist^b(A,B) \leqslant dist^b(A,C) + dist^b(B,C)$, that $dist^b(A,B) \leqslant \mathbf{R} - \mathbf{v}\Delta_{\text{relay}}$. Note that the relative locations and thus the distance of $A$ and $B$ are not controlled by the adversary. This implies that the adversary can violate ND1, only if the distance between $A$ and $B$ is smaller than $\mathbf{R} - \mathbf{v}\Delta_{\text{relay}}$ and $C$ is conveniently located.

On the other hand, for $\mathcal{A}'_{\Delta_{\text{relay}}}$ and setting $\mathcal{S}^c$ in Figure 2, $dist^c(A,C) + dist^c(D,B) + \frac{\mathbf{v}}{\mathbf{v}_{\text{adv}}} dist^c(C,D) + \mathbf{v}\Delta_{\text{relay}} \leqslant \mathbf{R}$. Utilizing this and the triangular inequality twice, that is, $dist^c(A,B) \leqslant dist^c(A,C) + dist^c(C,D) + dist^c(D,B)$, we get $dist^c(A,B) \leqslant \frac{\mathbf{v}_{\text{adv}}}{\mathbf{v}}(\mathbf{R} - \mathbf{v}\Delta_{\text{relay}})$. If the last inequality holds, the adversary can succeed with the use of an adversarial channel and two nodes $C, D$. It is interesting that the bound on $dist^c(A,B)$ is multiplied by a factor of $\frac{\mathbf{v}_{\text{adv}}}{\mathbf{v}}$. In other words, if $\mathbf{v} \ll \mathbf{v}_{\text{adv}}$, as it holds, for example, for ultrasound and radio frequency velocities [25], the use of the adversarial channel magnifies the impact on ND: the adversary can mislead nodes at remote locations (thus unable to communicate directly) that they are neighbors. Thus, whenever possible, the system designer should aim at having $\mathbf{v} = c$, which she can expect to be the choice of the adversary. This is further strengthened by the fact that the $\mathcal{P}^{\text{GT}}$ can be proven correct only if $\mathbf{v} = \mathbf{v}_{\text{adv}}$.

*Relation among Adversary Models.*

Intuitively, adversary $\mathcal{A}_2$ is stronger than adversary $\mathcal{A}_1$, if $\mathcal{A}_2$ can do everything that $\mathcal{A}_1$ can. Formally, this is expressed as follows:

DEFINITION 20. *Adversary model $\mathcal{A}_1$ is weaker[4] than adversary model $\mathcal{A}_2$ ($\mathcal{A}_1 \leqslant \mathcal{A}_2$), if $\Theta_{\mathcal{S},\mathcal{P},\mathcal{A}_1} \subseteq \Theta_{\mathcal{S},\mathcal{P},\mathcal{A}_2}$ for every setting $\mathcal{S}$ and every protocol model $\mathcal{P}$.*

Given this definition, we can order the considered adversary models:

$$\begin{array}{c} \mathcal{A}'_{\Delta_{\text{relay}}} \preceq \\ \mathcal{A}''_{\Delta_{\text{relay}}} \leqslant \end{array}^{5} \mathcal{A}_{\Delta_{\text{relay}}} \leqslant \mathcal{A}^{\mathsf{T}}_{\Delta_{\text{relay}}}$$

The relation among adversary models is interesting because one can intuitively expect that if a protocol $\mathcal{P}$ can solve ND for $\mathcal{A}_1$, it can also solve ND for a weaker adversary model $\mathcal{A}_2$.[6] Thus, our impossibility result, proven for the minimal elements, and the proof of correctness of protocol $\mathcal{P}^{\mathsf{T}}$ for the maximal element, hold for all adversary models considered in this paper. This clarifies that $\Delta_{\text{relay}}$ is the most significant factor affecting the security of ND, as opposed to the ability to use directional antennas, the adversary channel, or to generate arbitrary messages (in a Dolev-Yao fashion).

---

[4] non-strictly
[5] We use a different notation, $\mathcal{A}'_{\Delta_{\text{relay}}} \preceq \mathcal{A}_{\Delta_{\text{relay}}}$, as the "$\leqslant$" relation does not hold: in one case the adversarial nodes can only use Bcast and in the other only Dcast. However, Bcast($m$) is equivalent to a Dcast($0, 2\pi, m$). Accordingly, we can define a renaming function $\rho$, and show that the $\leqslant$ relation holds up to renaming: $\rho(\Theta_{\mathcal{S},\mathcal{P},\mathcal{A}'_{\Delta_{\text{relay}}}}) \subseteq \Theta_{\mathcal{S},\mathcal{P},\mathcal{A}_{\Delta_{\text{relay}}}}$.
[6] This can be proven under the assumption that the adversary model allows the adversarial nodes to remain silent, which is the case for all the adversary models that we consider. There exist adversary models for which this does not hold, but they are of no practical importance.

## 6.2 Modeling assumptions

Our ND specification and assumptions about wireless communication, protocols, and adversarial behavior all aim at a simple model. Nonetheless, these assumptions do not impair the generality and meaningfulness of our results. The discussion below establishes this mostly with respect to the impossibility result, as it is easy to see that most of these simplifying assumptions do not affect the ND protocols we model and prove correct.

*Protocol Model.*

Recall that our definition of a protocol model only requires that the behavior of the protocol is determined by the local view. This is much broader than the typical approach, in which a protocol is modeled by a Turing machine. But as our definition is an over-approximation, our impossibility result remains valid for more realistic protocol models.

*Settings and Traces.*

We emphasize that the general forms of settings (correct nodes being able to communicate at arbitrary distances), and Medium Access Control modeling (Definition 4 not prohibiting a correct node from sending and receiving an arbitrary number of messages at the same time) is *not* essential to the impossibility result. It is possible to add additional constraints to make the model more realistic, but this would impair generality and clarity.

*Events.*

We model correct nodes equipped with omnidirectional antennas. We can extend our model so that correct nodes use directional antennas, but from the structure of the impossibility result proof it should be clear that this would not lift the impossibility. Mounting a successful relay attack, however, would require adversarial node(s) to be located on or close to the line connecting $A$ and $B$.

We model success and failure (in fact, complete unawareness of failure) in receiving a message, but not the ability of a receiver to detect a transmission (wireless medium activity) without successfully decoding the message. An extension of our model to include this is straightforward and would not affect the impossibility result. Intuitively, if nodes were able to solve the ND problem if they cannot decode all the messages they receive, then they would also be able to solve ND when all messages are received correctly. We emphasize that the above argument relies on the assumption that nodes cannot control their wireless transmission power. However, if nodes had this ability, the notion of neighborhood would change, and our model would need to change as well. We will investigate this in future work.

*ND Specification.*

In light of the impossibility result, one could consider an alternative, less restrictive neighbor discovery specification, notably, the already mentioned *multi-party* ND that requires the participation of more than two nodes to securely conclude on a neighbor relation. This is an interesting direction resonating with emergent properties of ad-hoc networks [9]. Technically, this ND specification would differ in the ND2 property, where the requirement that the protocol needs to work for some *two-node* setting would be changed to an *arbitrary* setting. As discuss in Section 7, there exist protocols in the literature related to our notion of multi-party ND, but they are effective under weaker adversary models. Whether some other T-protocol can solve multi-party ND in our model is an open question we plan to investigate in future work.

*Line-of-sight Propagation.*

Definition 4 implies signal propagation over a straight line. In reality, this is not always the case, as two nodes could communicate even if there is no line-of-sight between them, and the signal is, for example, reflected. We could include this phenomenon in our model, for example, by introducing an additional link-specific delay to the propagation time. This would not affect any of our results. However, from a practical point of view, for such additionally delayed links, $\mathcal{P}^{\mathsf{T}}$ and especially $\mathcal{P}^{\mathsf{GT}}$ could reject valid neighbor relations. This problem relates to the discussion on inaccuracies in time and location information these protocols need to cope with in practice, in Section 6.3.

## 6.3 Protocol Design

We discuss some of the more important aspects for actual deployment of secure neighbor discovery protocols. First, we consider one side of ND: $A$ discovers if $B$ is a neighbor. However, with asymmetric links, a dual problem exists: $A$ discovers if it is a neighbor to $B$. The protocols we consider are not designed to solve this problem, but we note that challenge-response schemes, such as distance bounding protocols [2], can.

Moreover, we consider ND when both nodes running the ND protocol are correct. Removing this assumption implies that, for example, the $\mathcal{P}^{\mathsf{T}}$ protocol does not satisfy the ND specification: consider an adversarial node $B$ that generates a message time-stamped in the future, passes this message to another adversarial node $C$, which in turn passes it to a correct node $A$ that falsely accepts (a perhaps very remote) $B$ as a neighbor. In Section 7 two protocols that solve this problem under a specific assumption are discussed.

As mobility was not included in our model, the protocols we analyze can be considered secure as long as the node movement during the protocol execution is negligible. This is not a strong requirement, if we compare the typical speed at which nodes move (below the speed of sound in almost all cases) with the RF propagation speed. However, notably because some computational operations may be time-consuming, we plan to include mobility in our model in the future.

All the adversary models in this paper capture the technically feasible yet non-trivial ability to send and receive messages at the same time. For a weaker security result, one could assume that an adversarial node must receive the whole message before it can relay it. For such an adversary, a protocol whose every messages duration is longer than $\frac{R}{v}$ would solve ND (by Theorem 3).

Similarly to the vision of the authors of [13], $\mathcal{P}^{\mathsf{T}}$ and $\mathcal{P}^{\mathsf{GT}}$ functionality could be integrated into every packet as a leash. Alternatively, ND beacons can be broadcasted periodically, with the neighbor relation interpolated in between received beacons. The former solution provides better security at the expense of transmission overhead, whereas the latter might offer the adversary a window of opportunity to launch an attack if and only if the state of neighbor relation changes between two beacon broadcasts.

*Imperfect Clocks and Localization.*

Up to this point, we assumed that correct nodes have accurate time and location information. However, inaccuracies are possible in reality: (i) time inaccuracies due to clock drifts, failure to synchronize clocks, coarse-grained clocks, as well as the difficulty to calculate message reception time, and (ii) location inaccuracies due to unavailability of infrastructure (e.g., Global Positioning System (GPS), or base stations) providing location information, malicious disruptions of infrastructure, and granularity and capabilities of self-localization sensors. Non-line-of-sight propagation can be perceived as another source of time inaccuracy. As the $\mathcal{P}^{\mathsf{T}}$ and $\mathcal{P}^{\mathsf{GT}}$ protocols rely on distance estimates based on time and location measurements, their effectiveness can be affected by inaccuracies.

We model the effect of time inaccuracy by a parameter $\delta$, such that *measured delay = real delay + d*, with $|d| \leqslant \delta$. Similarly, for location information, *measured distance = real distance + s***v**, with $|s| \leqslant \tau$. We express the inaccuracy term $s\mathbf{v}$ as a function of delay (time), so that it is straightforward to consider the cumulative impact for the $\mathcal{P}^{\mathsf{GT}}$ protocol.

First, for $\mathcal{P}^{\mathsf{T}}$, two correct neighbors at a distance larger than $\mathbf{R} - \mathbf{v}\delta$ may fail to conclude they are neighbors, thus violating ND2. This can be addressed if $\mathbf{R}' = \mathbf{R} + \mathbf{v}\delta$ is used in place of the ND range $\mathbf{R}$. But then, if $\Delta_{\mathrm{relay}} < \frac{\mathbf{R}}{\mathbf{v}} + \delta$, or $\Delta_{\mathrm{relay}} < \frac{\mathbf{R}'}{\mathbf{v}}$, ND1 would be violated, that is, the adversary would mount a successful attack. In other words, time inaccuracies essentially decrease the ND security.

To cope with inaccuracies, the $\mathcal{P}^{\mathsf{GT}}$ protocol presented in Section 5 needs to be modified slightly: The check for *equality* of the time- and location-based estimates of distance should be replaced with *approximate equality*; otherwise ND2 will be violated. More precisely, these two estimates should be within $\delta + \tau$ of each other. But, again, ensuring practicality decreases security: if $\Delta_{\mathrm{relay}} < 2(\delta + \tau)$, the adversary could violate ND1.

More generally, for T-protocols, no additional consideration with respect to the impossibility results is necessary, as $\mathbf{R} \leqslant \mathbf{R}'$. But for TL-protocols, the inaccuracies in time and location could be viewed as an impossibility factor: for given $\delta$, $\tau$, there is no protocol solving the ND problem if the adversary can relay with delay $\Delta_{\mathrm{relay}} < 2(\delta + \tau)$. We emphasize however that the nature of these impossibility results differs, as it is not fundamental, as in the T-protocol case, but can be mitigated by introducing more sophisticated technology and obtaining accurate time and location, as long as line-of-sight propagation is assumed.

Finally, we note that accurate time and location information are not possible to achieve without specialized hardware. In addition, tight synchronization is nontrivial, but challenge-response protocols that do not need synchronized clocks can overcome this problem.

## 7. RELATED WORK

The prevalent wormhole prevention mechanism is based on *distance bounding*, which was first proposed by Brands and Chaum in [2] to thwart a relay attack between two correct nodes, also termed *mafia fraud*. Essentially, distance bounding estimates the distance between two nodes, with the guarantee that it is not smaller from their real distance. Subsequent proposals contributed in aspects such as mutual authentication [27], efficiency [10], and resistance to execution of the protocol with a colluding group of adversarial nodes [3, 24]. In the latter, the attack termed *terrorist fraud* is thwarted under the assumption that adversarial nodes do not expose their private cryptographic material; if not, one adversarial node can undetectably impersonate another and successfully stage a terrorist fraud. *Authenticated ranging*, proposed by Čapkun and Hubaux in [28], lifts the technically non-trivial requirement of rapid response (present in all the above protocols), at the expense of not being resilient to a *distance fraud*, when the protocol is executed with a *single, non-colluding* adversarial node [3]. This group of protocols, in which temporal packet leashes [13] and TrueLink [8] (both not resistent to the distance fraud) can be included, was the main inspiration for our investigation that led to a general impossibility result.

Another group of ND mechanisms is based on location, with geographical packet leashes [13] the primary representative. The impossibility result does not apply here, as T-protocols are not location-aware. Indeed, we prove that $\mathcal{P}^{\mathsf{GT}}$, an TL-protocol, *can* solve ND. We emphasize that $\mathcal{P}^{\mathsf{GT}}$ is different from geographical packet leashes, because it requires clock synchronization as tight as that for temporal packet leashes. Essentially, $\mathcal{P}^{\mathsf{GT}}$ is a combination of temporal and geographical leashes. Upon careful inspection of the literature, there exist prior passages seemingly cluing or relating to this idea: the introduction of [12] or the discussion of combining a so-called node-centric localization scheme with distance bounding techniques [29]. Nonetheless, to the best of our knowledge, we are the first to explicitly point out the advantages, over other approaches for secure ND, of combining location information with tight temporal bounds. We note that the authors of [13] mention the obstacle problem, but only in the case of geographical packet leashes. However, the solution that they propose – having a radio propagation model at every node – is not applicable in most scenarios.

The approach of Poovendran and Lazos [21] can be seen as an extension of a location based scheme: a few trusted nodes (*guards*) are aware of their location, transmit it periodically in beacons, and all other nodes determine their neighbors based on whether they received sufficiently many common beacons. This scheme is a multi-party ND protocol and thus our impossibility result does not apply. Unfortunately, from the perspective of our approach, [21] has some serious drawbacks. Most notably, it relies on the "no obstructions" assumption – nodes that are close but cannot communicate can be tricked into establishing a neighbor relation. In addition, adversarial nodes are rather limited in their behavior: one can see an attack against this scheme, in particular Claim 2, when adversarial nodes are allowed to selectively relay beacon messages.

A scheme using directional antennas was proposed by Hu and Evans in [12], with the interesting property that it can be used as a two-party ND protocol, or as a multi-party ND protocol with additional nodes serving as *verifiers* of neighbor relations. In the two-party operation the scheme has security weaknesses that the multi-party version is called upon to remedy. In the latter case, our impossibility result does not apply directly. Nonetheless, significant security problems remain, with the scheme oblivious to obstacles and the adversary model limited. As the authors point out, a successful attack can be mounted if more than two adversarial nodes collaborate. Recall that in our proofs we allow for arbitrary node collaboration (or collusion).

[14] proposes to collect local, $k$-hop connectivity information obtained with a non-secure ND mechanism, and to inspect it for *forbidden structures*: subgraphs that are likely to exist only if a wormhole is present in the vicinity. The exchange of connectivity information makes it a multi-party protocol. Although the simulations presented in [14] show a very good detection rate, as in [21], the considered adversary is quite naive: a single non-selective long-range wormhole.

A different approach to secure neighbor discovery could exploit radio frequency fingerprinting (RFF) [4]: devices from the same production line are not identical, but rather the signals each one emits may have unique identifiable features. If these signals can be identified upon reception of a message, it becomes impossible for an adversarial node to relay any message undetected. If such a scheme were in place, our impossibility result would not apply. The reason is that impossibility hinges on the very fact that a correct node cannot identify how a message was received. This essentially allows the adversary to relay wireless transmissions (messages). However, it is questionable if RFF can be used to secure ND. Investigations with different types of devices, e.g., [23] or [26], show classification success rate around 90% in laboratory conditions. At the same time, findings such as "... radios were found to have fingerprints that were virtually indistinguishable from each other, making the identification process more difficult, if not impossible..." [7] clue on unresolved limitations.

The wormhole attack, in its symptoms, bears similarity to two other fundamental and hard to detect attacks. On one hand, a wormhole end can be perceived as a *Sybil* node, with messages tied to different identities being transmitted by a single node. Hence, seemingly, a Sybil node detection mechanism [17] could be used to thwart relay attacks. However, a wormhole can selectively relay the messages of a single node, and still be effective (e.g. Figure 2, setting $\mathcal{S}^c$). On the other hand, as in the *node replication* attack, messages tied to a single identity are transmitted by more than one node. However, node replication is harder to detect than a wormhole attack: schemes that address node replication [20, 5] focus on probabilistically detecting replicas located in remote parts of the network and require that nodes are location-aware. Obviously, a long-range wormhole can be easily (and deterministically) prevented using geographical packet leashes.

A large body of work on formal reasoning on cryptographic protocols exists, yet the classical cryptographic protocols live in the Internet: thus these methods are agnostic about the characteristics of the communication medium, especially a wireless one. Recently, there has been a rising interest in formalizing analysis of security protocols in wireless networks. The problem of distance bounding has been treated formally in [15], whereas other works were concerned with routing [16, 1, 18, 30] or local area networking [11]. These works are concerned with different problems and their approaches are not amenable to reason about secure neighbor discovery.

## 8. CONCLUSIONS

We investigate the problem of secure neighbor discovery (ND) in wireless networks. We build a formal framework, and provide a specification of neighbor discovery or, more precisely, its most basic variant: two-party ND. We consider two general classes of protocols: time-based protocols (T-protocols) and time- and location-based protocols (TL-protocols). For the T-protocol class, we identify a fundamental limitation governed by a threshold value depending on the ND range: We prove that no T-protocol can solve the ND problem if and only if adversarial nodes can relay messages faster than this threshold. This result is a useful measure of the ND security achieved by T-protocols and leads us to investigate other classes of protocols.

In particular, we prove that no such limitation exists for the class of TL-protocols: They can solve the ND problem for any adversary, as long as the time and location measurements are accurate enough, and line-of-sight signal propagation is assumed. The protocols we analyze are very simple if not the simplest possible to allow positive results. In future work, we will focus on a larger spectrum of protocols, most notably multi-party neighbor discovery, as well as model additional aspects, such as the ability of nodes of controlling their transmission power.

## 9. ACKNOWLEDGMENTS

The authors would like to thank Patrick Schaller, David Basin, Srdjan Čapkun, Pascal Lafourcade and Paul Hankes Drielsma for the inspiring discussions and the anonymous reviewers for their helpful comments.